\documentclass[fleqn,usenatbib]{mnras}

\usepackage{newtxtext,newtxmath}

\usepackage[T1]{fontenc}

\DeclareRobustCommand{\VAN}[3]{#2}
\let\VANthebibliography\thebibliography
\def\thebibliography{\DeclareRobustCommand{\VAN}[3]{##3}\VANthebibliography}

\usepackage{graphicx}	
\usepackage{amsmath}	

\def\sm1606{SMSS~J1606$-$1000}

\title[The magnetic system SMSS~J1606-1000]{The  magnetic system SMSS~J1606-1000 as a period bouncer}

\author[A. Kawka, et al.]{Adela Kawka,$^{1}$\thanks{E-mail: adela.kawka@curtin.edu.au (AK)}
St\'ephane Vennes,$^{2}$ Lilia Ferrario,$^{2}$
M.S. Bessell,$^{3}$
S.C. Keller,$^{3}$
E. Paunzen,$^{4}$
\newauthor 
D.A.H. Buckley,$^{5,6}$
D. Groenewald,$^{5,7}$
J. Jan\'{\i}k$^{4}$
and M. Zejda$^{4}$
\\
$^{1}$International Centre for Radio Astronomy Research - Curtin University, GPO Box U1987, Perth, WA 6845, Australia \\
$^{2}$Mathematical Sciences Institute, The Australian National University, ACT 0200, Australia\\
$^{3}$Research School of Astronomy and Astrophysics, The Australian National University, Canberra, ACT 2611, Australia \\
$^{4}$Department of Theoretical Physics and Astrophysics, Masaryk University, Kotl\'a\v{r}sk\'a 2, CZ-611 37, Czech Republic\\
$^{5}$South African Astronomical Observatory, Observatory Road, Observatory 7935, South Africa\\
$^{6}$Department of Astronomy, University of Cape Town, Rondebosch 7770, Cape Town, South Africa\\
$^{7}$South African Large Telescope, PO Box 9, Observatory 7935, South Africa\\
}

\date{Accepted XXX. Received YYY; in original form ZZZ}

\pubyear{2021}

\begin{document}
\label{firstpage}
\pagerange{\pageref{firstpage}--\pageref{lastpage}}
\maketitle

\begin{abstract}
We report the discovery of a rare close binary system, SMSS~J160639.78-100010.7, comprised of a magnetic white dwarf with a field of about 30~MG and a brown dwarf. We measured an orbital period of 92\,min which is consistent with the photometric period. Minimum and maximum light occur at the orbital quadratures $\Phi=0.25$ and 0.75, respectively, and cannot be caused by reflection on the brown dwarf, but, instead, by a spot on the synchronously rotating magnetic white dwarf. The brown dwarf does not fill its Roche lobe and the system may be in a low-accretion state or, more likely, in a detached state following episodes of mass transfer. SMSS~J160639.78-100010.7 is the nearest known magnetic white dwarf plus brown dwarf system.
\end{abstract}




\section{Introduction}

White dwarfs are often found in close binaries with low mass main-sequence companions (usually M dwarf) although their number drops sharply for companions later than an M5 type \citep{fer2012} and only a handful of white dwarfs has been found in either interacting or detached binaries containing a brown dwarf. Such systems can be detected through infrared excess and recent surveys have identified several candidates \citep[e.g.,][]{ste2011,gir2011,hog2020}. \citet{ste2011} estimated that only 0.5 per cent of white dwarfs have a brown dwarf companion.

Only a few binary systems comprising a magnetic white dwarf and a close but apparently detached low mass companion are known.
Initially, these systems were thought to be low accretion rate polars \citep[LARPs;][]{rei2000}, until \citet{Tout2008} suggested that, hidden among LARPs, there could be systems known as the pre-polars or PREPs \citep{sch2009}, that emerged from common envelope evolution as close pairs but still await for gravitational radiation to bring them close enough for mass transfer to take place.

In this paper we introduce SMSS~J160639.78-100010.7 (hereafter \sm1606) which is a new highly magnetic white dwarf with a brown dwarf companion in a close orbit. This system is part of a selection of blue objects extracted from the SkyMapper survey \citep{onk2019}. New optical spectroscopic and photometric data together with archival infrared and ultraviolet data are presented in Section 2 followed by our analysis of the binary parameters in Section 3.1, the photometric variations in Section 3.2, the character of the individual components of the system in Section 3.3, and of the system kinematics in Section 3.4. A discussion and our conclusions follow in Section 4.

\section{Observations}

We first obtained a spectrum of \sm1606\ with the FORS2 spectrograph attached to the Very Large Telescope at the European Southern Observatory on UT 2016 March 11. We used the 600B grism and set the slit width to 1 arcsec to provide a spectral coverage from 3500 to 6000\AA\ at a resolution of $\approx 6$ \AA. The spectrum revealed a Zeeman-split hydrogen line spectrum in a strongly magnetic white dwarf with H$\beta$ and weaker H$\gamma$ emission lines. The emission lines do not show the effect of a magnetic field.

We followed-up with a series of spectra obtained with the Wide-Field Spectrograph \citep[WiFeS:][]{dop2010} attached to the 2.3~m telescope
at Siding Spring Observatory (SSO) on UT 2016 April 8, 2020 May 27 to 29, 2021 February 20, April 5, May 3, and June 19. We used the B3000 and R3000 gratings which
cover a combined spectral range from 3500 to 9000 \AA\ during the 2016 April and 2020 May runs, and the B3000 and R7000 gratings with a coverage from 3500 to 7000\AA\ during all subsequent runs. We set the exposure times to 1800~s in the 12 exposures obtained on 2016 April 8 and 2020 May 27 and reduced the exposure times to 1200~s in all subsequent 48 exposures. Frequent NeAr arc spectra were obtained each night. The data were reduced using the PyWiFeS reduction pipeline \citep{chi2014}. The pipeline generates three-dimensional data sets consisting of spatially resolved wavelength and flux calibrated spectra for each slitlet. We then extracted background-subtracted spectra from the slitlets that provided significant flux from \sm1606.

We obtained a photometric time series of \sm1606\ with the 1.54~m telescope at La Silla on UT 2018 April 6. We used the $R$ filter and the series lasted
$\approx 4.8$~hr with 125 exposures of 120~s each.

We collected ultraviolet photometric measurements from the Galaxy Evolution Explorer \citep[GALEX:][]{mor2007}, optical photometric 
measurements from SkyMapper \citep{onk2019} and infrared photometric measurements from the VISTA Hemisphere Survey \citep[VHS:][]{mcm2013}.
These are listed in Table~\ref{tbl_astrometry} together with photometric and astrometric measurements obtained by {\it Gaia} \citep{gai2020}.

\sm1606\ is not a known X-ray source.

\begin{table}
 \caption{Astrometry and photometry of \sm1606.}
 \label{tbl_astrometry}
 \begin{tabular}{lcclcc}
  \hline
  \multicolumn{2}{l}{Parameter} & \multicolumn{3}{c}{Measurement} & Ref. \\
  \hline
  \multicolumn{2}{l}{RA (J2000)} & \multicolumn{3}{c}{16$^{\rm h}$06$^{\rm m}$39\fs78} & 1 \\
  \multicolumn{2}{l}{Dec (J2000)} & \multicolumn{3}{c}{$-$10\degr 00\arcmin 10\farcs7} & 1 \\
  \multicolumn{2}{l}{$\mu_\sigma \cos{\delta}$ (\arcsec yr$^{-1}$)}& \multicolumn{3}{c}{$-0.1446\pm0.0002$} & 2 \\
  \multicolumn{2}{l}{$\mu_\delta$ (\arcsec yr$^{-1}$)} & \multicolumn{3}{c}{$-0.0796\pm0.0001$} & 2 \\
  \multicolumn{2}{l}{$\pi$ (mas)} & \multicolumn{3}{c}{$9.27\pm0.13$} & 2 \\
\hline
\multicolumn{6}{c}{Photometry} \\
\hline
Band & Measurement & Ref. & Band & Measurement & Ref. \\
\hline
  $G$   & $17.944\pm0.003$ & 2 & $g$ & $17.993\pm0.067$ & 1 \\
  $b_p$ & $17.969\pm0.010$ & 2 & $i$ & $18.088\pm0.010$ & 1 \\
  $r_p$ & $17.754\pm0.025$ & 2 & $z$ & $18.138\pm0.063$ & 1 \\
  $FUV$ & $20.260\pm0.210$ & 3 & $Y$ & $17.895\pm0.034$ & 4 \\
  $NUV$ & $18.841\pm0.047$ & 3 & $J$ & $17.387\pm0.033$ & 4 \\
  $u$ & $18.259\pm0.025$ & 1 & $H$ & $15.874\pm0.018$ & 4 \\
  $v$ & $18.145\pm0.028$ & 1 & $K$ & $15.243\pm0.022$ & 4 \\
  \hline
 \end{tabular}\\
References: (1) \citet{onk2019}; (2) \citet{gai2020}; (3) \citet{mor2007}; (4) \citet{mcm2013}. 
\end{table}

\section{Analysis}

Prompted by the presence of Balmer emission lines in the FORS2 spectrum presumably emerging from or near the upper atmosphere of a cool companion, we proceed with a period analysis of the radial velocity and photometric measurements. Next, we present an analysis of the spectroscopic and photometric measurements of the system revealing an infrared excess attributed in part to the cool companion.

\subsection{Binary properties}

\begin{figure}
    \centering
    \includegraphics[viewport=70 158 500 790,clip,width=0.80\columnwidth]{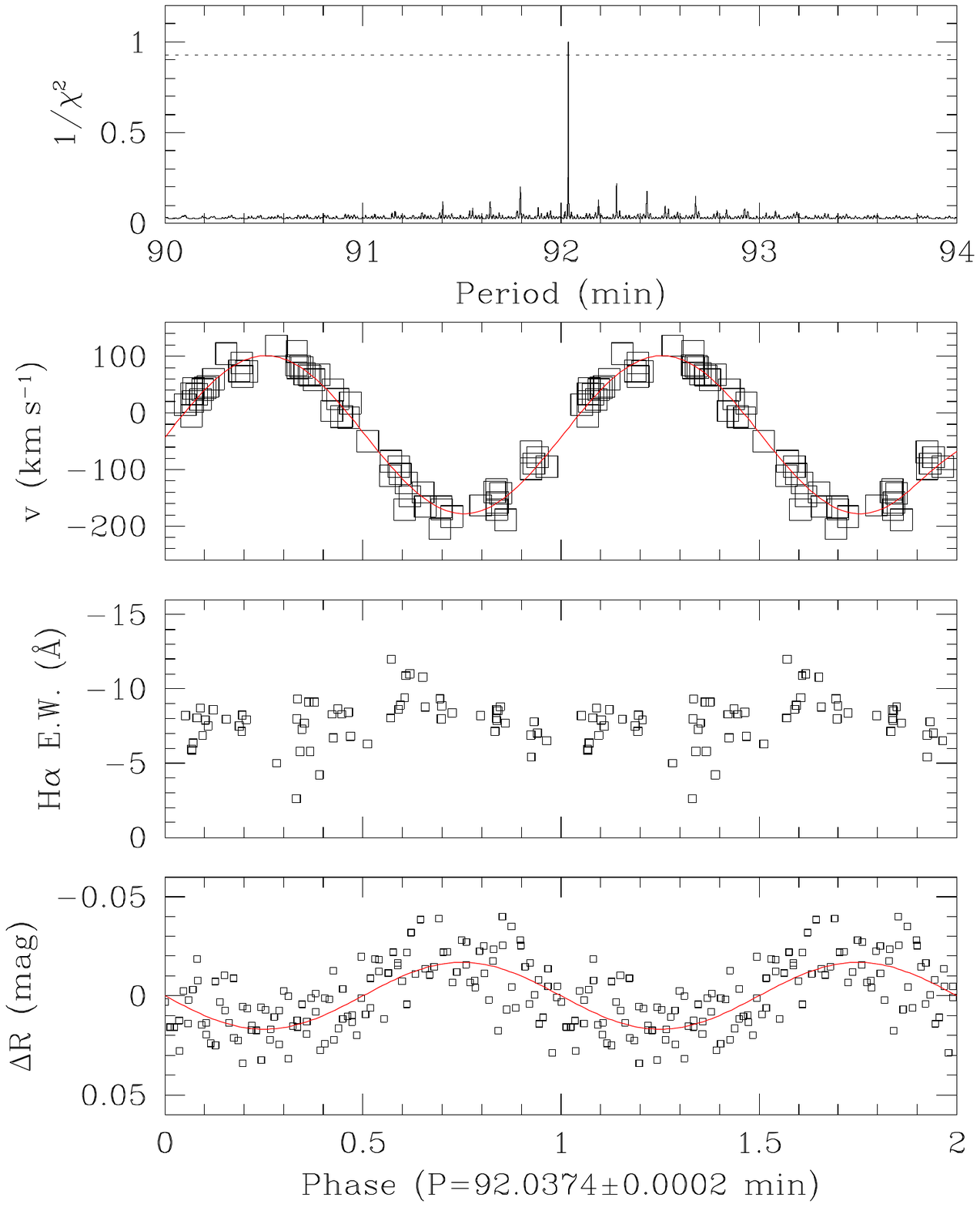}
    \caption{Top: Periodogram of the WiFeS H$\alpha$ radial velocity measurements with the 90 percent confidence level shown with a dashed line. The phased radial velocity (second panel from top), H$\alpha$ equivalent width (third panel), and R-band photometric measurements (bottom panel) are shown phased with the binary ephemeris.}
    \label{fig_vel}
\end{figure}

\begin{figure}
    \centering
    \includegraphics[viewport=58 185 348 600,clip,width=0.75\columnwidth]{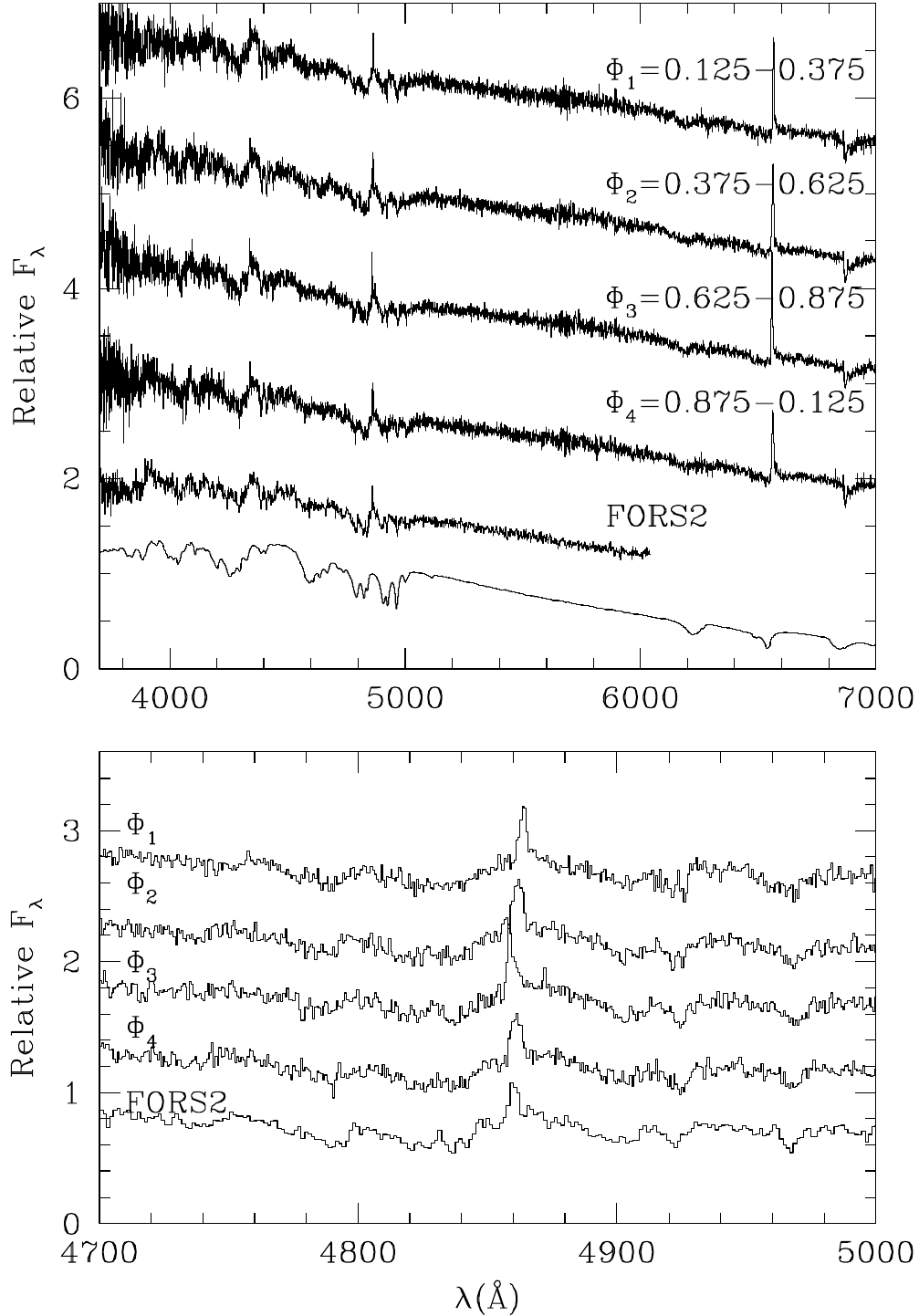}
    \caption{Top: WiFeS flux spectra binned at four consecutive orbital phases and FORS2 spectrum compared to a model spectrum (unlabeled) at $T_{\rm eff} = 9600$~K and
    $\log{g} = 8.2$ with a dipolar field of 30~MG (see Section 3.2). The Balmer emission spectrum at H$\alpha$, H$\beta$ and H$\gamma$ is in evidence on top of the white dwarf absorption spectrum. Bottom: Same spectra but zoomed in around H$\beta$ showing variability between the four phases.}
    \label{fig_wifes}
\end{figure}

We measured the radial velocities using the H$\alpha$ emission lines. The velocities were corrected for the solar system's barycentre motion and we fitted these with a sinusoidal function
\begin{displaymath}
v(t) = \gamma + K\,sin[2\pi(t-T_0)/P]
\end{displaymath}
where $P$ is the orbital period, $\gamma$ is the systemic velocity, $K$ is the velocity semi-amplitude, and $T_0$ is the initial epoch of inferior conjunction of the cool companion. Fig.~\ref{fig_vel} shows a period analysis of the radial velocities showing that the orbital period is $P=92.0374\pm0.0002$~min. The velocity semi-amplitude is 141.2$\pm$3.2~km~s$^{-1}$ and the systemic velocity $-$38.4$\pm$2.2~km~s$^{-1}$. However, the velocity measurements suffer from orbital smearing because the exposure times correspond to approximately one quarter of the orbital period. We corrected for this effect by applying boxcar smoothing (full-width of 0.24 phase corresponding to the average exposure time divided by the orbital period) to the sinusoidal functions and by refitting the velocity measurements. We obtained a corrected velocity amplitude of 153.7$\pm$3.5~km~s$^{-1}$. A correction for surface variations of the Balmer emission was not included but it may amount to $\approx$8 per cent if the Balmer emission is restricted to the hemisphere facing the white dwarf.  Fig.~\ref{fig_wifes} shows the WiFeS spectra averaged over four orbital phase bins. The H$\beta$ emission traces the brown dwarf motion and absorption features trace the white dwarf motion. By cross-correlating the H$\beta$ absorption features from 4700 to 4830 and from 4890 to 5100\AA\ in the four phased spectra we find a white dwarf velocity semi-amplitude of $-$38.9$\pm$27.5~km~s$^{-1}$ and an uncertain binary mass ratio $q=M_{\rm BD}/M_{\rm WD}=0.08$-0.5 which is only consistent with our spectroscopic analysis at the lower end, i.e., with a brown dwarf ($q=0.08$-0.10, see section 3.3). The white dwarf orbit is poorly constrained by current data as the line profiles are possibly affected by surface field variations. 

\subsection{Photometric variations}

The $R$ band photometric measurements show variations with a semi-amplitude $\Delta m=0.017\pm0.002$ mag and a period of $93.3\pm5.9$~min (Fig.~\ref{fig_vel}, bottom panel). 
The photometric period is consistent with the orbital period. Maximum reprocessed light in a close pair involving a hot compact star with a cool tidally-locked companion should occur at the inferior conjunction of the hot primary ($\Phi$=0.5). However,
we measured maximum light at orbital quadrature $\Phi$=0.75 with a phase offset of only $-0.0007\pm0.0011$ implying
that the photometric variations are probably due to a magnetic spot on the white dwarf surface assuming synchronous rotation, i.e., that the rotational period of the white dwarf is identical to the orbital period. Ellipsoidal variations could produce a peak at $\Phi$=0.75, 
but it would also produce a peak at $\Phi=0.25$ that is not observed.
Photometric observations in multiple bands are needed to investigate further the source of the variations.

\sm1606\ was observed with GALEX over four separate visits in the $NUV$ and two visits in the $FUV$. 
We used {\sc gPhoton} \citep{mil2016} to extract $NUV$ magnitudes from 4 separate observations to check for 
possible variability. The $NUV$ measurements show  variability of up to 0.2 magnitudes but with only a 40 percent probability because of the large uncertainty in the individual measurements ($\sigma=$ 0.11 - 0.13). The $FUV$ observations are very uncertain and do not provide any constraints on variability.

\subsection{Atmospheric and stellar parameters}\label{atm}

\begin{figure}
    \centering
    \includegraphics[viewport=32 166 565 702,clip,width=0.85\columnwidth]{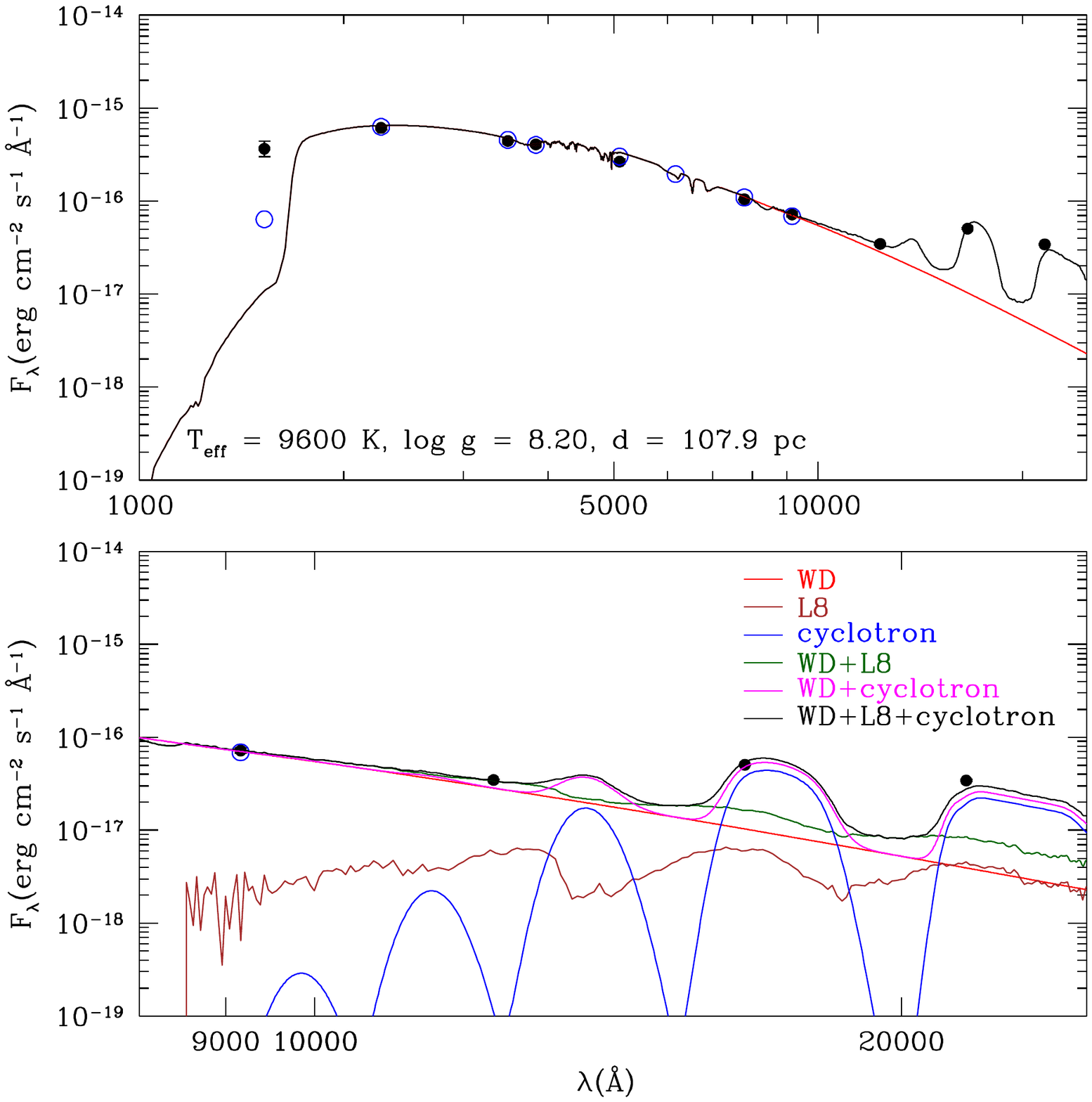}
    \caption{The spectral energy distribution of \sm1606. The top panel shows photometric measurements compared to a
    white dwarf model spectrum at an effective temperature of 9600~K and surface gravity $\log{g}=8.2$ (in red) and the combined spectrum
    which includes contribution from the white dwarf, brown dwarf and cyclotron emission. The bottom panel shows the various 
    flux contributions in the IR.}
    \label{fig_sed}
\end{figure}

We have used the spectral energy distribution (SED) combined with the distance measurement from the Gaia parallax to constrain the white dwarf and companion properties, while accounting for possible cyclotron emission.  
We modeled the magnetic white dwarf using a prescription for the field distribution from \citet{mar1984} and \citet{ach1989}. The visible surface was modeled using 450 elements co-added to provide a surface-averaged emergent spectrum. We utilized Stark broadened hydrogen line profiles with field-dependent line positions and strengths from \citet{sch2014}.
For completeness, we added Zeeman-shifted quasi-molecular Ly$\alpha$ opacities in the FUV spectral range with zero-field line opacities from \citet{all1992}.
The effective temperature and surface gravity
of the white dwarf are $9600\pm300$~K and $\log{g} = 8.2\pm0.08$, respectively, corresponding to a mass of $0.72\pm0.05\,M_\odot$ and, assuming no prior interactions, a cooling age of $9.7\pm0.6\times10^8$~yr \citep{ben1999}.
The white dwarf progenitor mass is between $\approx 2.5$~M$_\odot$ and $3.5$~M$_\odot$ 
\citep{ren2010} with a corresponding main-sequence lifetime ranging from $3\times 10^8$~yr to $7\times 10^8$~yr \citep{rom2015} 
for a total age of 1.3-1.6~Gyr which may fall short due to likely past interactions.
We modeled the spectral Zeeman components and found that we could
match the main features with a dipolar field of 30~MG at an inclination of $65^{\circ}$ with an offset along the magnetic axis of $+0.2$. Fig.~\ref{fig_wifes}
compares a model spectrum to four phase-binned WiFeS spectra obtained during the May 2020 run. The spectra show that the Zeeman-split Balmer lines vary as a function of the phase indicating that the magnetic field strength varies across the white dwarf surface.

Fig.~\ref{fig_sed} shows the SED of \sm1606\ comparing the
flux contribution from the white dwarf, the companion and the cyclotron emission. We scaled the flux from the brown dwarf template on the absolute $H$ magnitude from \citet{fah2012} set at a distance of 107.9~pc.
The SED  near the VISTA-$J$ band constrains the companion to be a brown dwarf with a spectral type of about L8 \citep{tes2001} which corresponds to an effective temperature of $\sim 1500$~K \citep{fil2015}. The $GALEX$/$FUV$ flux is significantly higher than the predicted flux from the white dwarf models suggesting the effect of UV excess due to a hot spot on the white dwarf.

\begin{figure}
    \centering
    \includegraphics[viewport=35 155 575 550,clip,width=0.95\columnwidth]{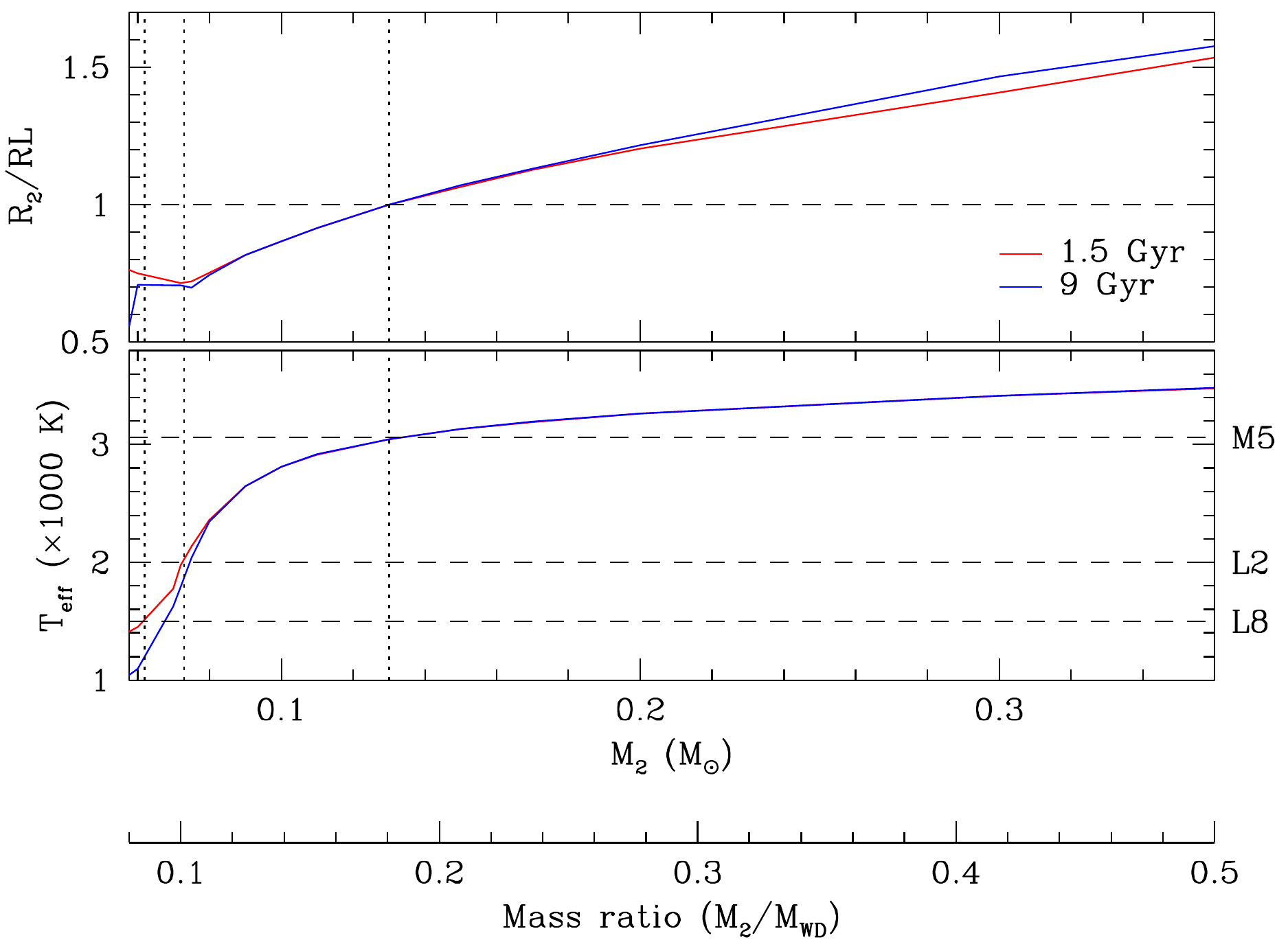}
    \caption{Top: Ratio of the secondary radius ($R_2$) to the Roche lobe radius ($RL$) at two different ages as a function of the secondary mass ($M_2$) or, alternatively, the binary mass ratio assuming $M_{\rm WD}=0.72\,M_\odot$. The rightmost vertical line marks a filled Roche lobe, and the middle and leftmost lines mark, respectively, the secondary mass upper limit and the preferred mass obtained from the SED analysis. Bottom: Corresponding effective temperatures ($T_{\rm eff}$) for the same relations depicted in the top panel showing the resulting constraints on the spectral type (horizontal lines).}
    \label{fig_rl}
\end{figure}

Fig.~\ref{fig_rl} shows available constraints on the secondary mass, radius and spectral type at an age of 1.5 Gyr and up to 9 Gyr \citep{bar2015}. The Roche lobe radius was calculated following \citet{egg1983}. We found that only a secondary star earlier than M5 would fill its Roche lobe. However, the SED analysis points toward an L8 brown dwarf secondary and no earlier than L2. Adopting an L8 type and accounting for a wide range of ages (1.5-9 Gyr), we estimate a secondary mass of 0.06-0.07~$M_\odot$ or a mass ratio $q=0.08$-0.1, and a secondary radius $R_2=$0.086-0.092~$R_\odot$.

The brown dwarf accounts for some of the IR excess in the SED, and the remaining excess can only be explained by a hot spot with a cyclotron
absorption spectrum. We have calculated the cyclotron spectrum following \citet{cha1981} and \citet{eng1973}. To match the IR photometric measurements we
used a thermal plasma of $kT = 2.0$~keV with a thickness of $9\times10^3$cm, an electron density of $10^{16}$~cm$^{-3}$, and a uniform magnetic field adjusted to 15.5~MG at an inclination of $75^\circ$. The cyclotron emission region covers 0.04 per cent of the white dwarf's surface and is located near the weaker but more favourably inclined pole at $\approx15$~MG.
Infrared spectroscopy is needed to disentangle the flux contribution from the brown dwarf and the cyclotron emission.

\subsection{Kinematics}\label{kin}

We calculated the Galactic velocity components relative to the local standard of rest using the
distance, proper motion and systemic velocity following \citet{joh1987}. The velocities of \sm1606\ were corrected for the Solar motion relative to the local standard of rest using ($U_\odot$,$V_\odot$,$W_\odot$)=(11.10, 12.24, 7.25)~km~s$^{-1}$ \citep{sch2010}. The Galactic velocity components of ($U$,$V$,$W$) = 
(-32,-68,16)~km~s$^{-1}$ place \sm1606\ in the thick disc \citep{sou2003}. Next, we 
calculated the eccentricity, $e=0.34$, and the $z$-component of the angular momentum, $L_z$ = 
1200~kpc~km~s$^{-1}$, of the Galactic orbit using \textsc{galpy} \citep{bov2015} where we used the \texttt{MWPotential2014} Galactic potential. These properties
place \sm1606\ in the thick disc following the classification of \citet{pau2006} and with an age of $\approx$9-10~Gyr \citep{sha2019} well in excess of the white dwarf age ($1.3$-$1.6\times10^{9}$ yr) derived assuming single star evolution. \sm1606\ must be an old system that experienced delays in the white dwarf cooling due to accretion.

\begin{table}
    \centering
    \label{tbl_param}    
    \caption{Atmospheric and orbital parameters of \sm1606.}
    \begin{tabular}{lc}
\hline
Parameter & Measurement \\
\hline
$T_{\rm eff ,WD}$ (K) & 9600$\pm$300 \\
$\log{g}$ (cgs) & $8.2\pm0.08$ \\
Mass (M$_\odot$) & $0.72\pm0.05$ \\
$B_{\rm WD}$ (MG) & 30 \\
Period (d) & $0.06391487\pm0.00000014$ \\
$T_0$ (HJD) & $2458436.2144\pm0.0015$ \\
$K_{\rm BD}$ (km s$^{-1}$) & $141.2\pm3.2$ \\
$K_{\rm BD,corr}$ (km s$^{-1}$) & $153.7\pm3.5$ \\
$\gamma$ (km s$^{-1}$) & $-38.4\pm2.2$ \\
\hline
\end{tabular}
\end{table}

\section{Discussion and Conclusions}

\sm1606\ belongs to a sub-class of close binaries consisting of a white dwarf paired with a very low mass companion. Such binaries exhibit a peak in the M dwarf companion distribution near spectral type M3.5 with a very steep decline at spectral types later than M5 \citep{fer2012}. The survey of \citet{far2005} found, relative to this peak,  $\approx2-3$ times more L dwarfs and $\approx 4-5$ times more M6–M9 dwarfs in the field than among companions to white dwarfs. Despite more recent surveys with greater sensitivity to late type M dwarfs and brown dwarfs, very few of these systems have been detected \citep[e.g][]{hog2020}. Furthermore, \citet{car2020} have reported that there are only 23 brown dwarfs transiting main-sequence stars with orbits within 10\,AU revealing that the `brown dwarf desert' puzzle still persists. This phenomenon observed by \citet{mar2000} consists of a lack of brown dwarfs within 3\,AU of a main-sequence star. The descendants of these binaries, a white dwarf with a brown dwarf companion, are even more infrequent. If the white dwarf is magnetic, such a pairing becomes very rare. Only six magnetic systems with very late M type or brown dwarf companions are known (Table \ref{tbl_MWD_BD}). 

The component masses derived in the SED analysis of \sm1606\ resulted in a binary mass ratio $q = 0.08$-0.1. 
We derived a Roche lobe radius $R_L=0.11$-0.13$~R_\odot$ for the brown dwarf. With a radius of only 
$0.086$-0.092$~R_\odot$ \citep{bar2015}, the brown dwarf would fill at most 60 per cent of its Roche lobe. 
The system \sm1606\ appears detached.

In the case of SDSS~J1212$+$0136, early studies suggested that the brown dwarf was under filling its Roche lobe and was more likely to be a PREP than a LARP \citep{sch2005a, far2008}. However, 
the X-ray observations of \citet{ste2017} strongly suggest that this system may be filling its Roche lobe and should be considered a LARP.

Other magnetic systems with a brown dwarf secondary are SDSS~J1514$+$0744 and SDSS~J1250$+$1549 \citep{bre2012}. The spectral type of the companion in IL\,Leo (=SDSS\,J1031$+$2028) \citep{sch2007, pars2021} has not been established yet. Finally, EF\,Eri was a persistent and bright X-ray source from its discovery in 1979 \citep{wil1979} until it entered a low accretion state in 1997 \citep{beu2000}; its companion is probably an L or T dwarf \citep{beu2000, Howell2006}. 

Systems with very low mass companions could either be (i) polars in a prolonged low-state of accretion, like EF\,Eri, or (ii) detached binaries that will eventually undergo Roche lobe overflow, or (iii) old polars that have reached the orbital period minimum $P_{\rm orb}$ for CVs and are evolving towards longer periods (period bouncers). Genuine PREPs, however, are expected to exhibit longer orbital periods \citep[$>2.5$\,hrs,][]{fer2015} and this seems to be the case for most systems currently classified as PREPs, except for IL\,Leo (=SDSS\,J1031$+$2028). IL\,Leo has $P_{\rm orb}=82$\,min, a white dwarf with a field of 42\,MG and a companion of spectral type later than M6 and likely to under fill its Roche lobe \citep{sch2007}.
 
If a binary is a PREP, the brown dwarf in the system should be in its original state. This hypothesis is supported by the observations of \citet{max2006} showing that the brown dwarf companion to the white dwarf WD\,0137$-$349 was engulfed by the red giant but survived the experience unscathed. 
 
However, if the brown dwarf belongs to a LARP or to a period bouncer, it was born as a hydrogen burning M dwarf that was peeled off by mass transfer leaving a degenerate sub-stellar object that can no longer burn fuel. Its structure and chemical composition (e.g., higher helium abundance and no lithium) would differ from those of a genuine brown dwarf. 

The evolution of CVs is driven by angular momentum losses from the orbit. Because of mass transfer, both the mass of the secondary star and the orbital period decrease during evolution, but when the secondary's thermal time-scale and the mass transfer time-scale
equal each other the period starts increasing. This period bounce occurs when $M_2\approx 0.07$\,M$_\odot$ \citep{pac1981}.

\begin{table}
\begin{minipage}{\columnwidth}
\centering
\caption{Magnetic systems with a late M or brown dwarf secondary star}
\label{tbl_MWD_BD}
\begin{tabular}{@{}l@{}rcrccc@{}c@{}}
\hline
 Name & P$_{\rm orb}$ & M$_{\rm WD}$ & T$_{\rm eff}$ & $B$  & type  & Ref. \\
      & (min) & ($M_\odot$) & (K) & (MG)  & & \\
    \hline
    1606$-$1000\,   & 92 & 0.72    &  9\,600 & 30 &  L8  & 1 \\
    1514$+$0744\,   & 89 & ...     & 10\,000 & 36 &  L3 & 2 \\
    1212$+$0136\,   & 88 & $>$0.41 & 9\,500  & 7-13 & L5-8 & 3,4,5\\
    1250$+$1549\,   & 86 & $>$0.42 & 10\,000 & 20 &  M8  & 2 \\
    IL\,Leo       & 82 & $>$0.48 &$<$11\,000&42 & $>$M6& 6,7 \\
    EF\,Eri       & 81 & $>$0.65 & 9\,850 & 16-21 & $>$L4&  8,9,10\\
    \hline
\end{tabular}\\
References: (1) This work; (2) \citet{bre2012}; (3) \citet{sch2005a}; (4) \citet{bur2006}; (5) \citet{far2008}; (6) \citet{pars2021}; 
(7) \citet{sch2007}; (8) \citet{fer1996}; (9) \citet{szk2010}; (10) \citet{beu2000}
\end{minipage}
\end{table}

\sm1606\ is only the third known weakly- or non-interacting magnetic white dwarf plus brown dwarf binary after SDSS~J1212+0136,
SDSS~J1514+0744. The properties of these systems are similar with orbital periods near 90 minutes and white dwarf effective temperatures near 10\,000~K. \sm1606\ has a longest
orbital period of the three. Based on the kinematics, we found that \sm1606\ is a very old system ($>9$~Gyr) in the thick disc which excludes a PREP classification. 
The relatively long period and low mass ratio suggest that \sm1606\ is a period bouncer rather than a LARP.

A possible evolutionary scenario involves a wide main-sequence binary with a 2.5~$M_\odot$ primary and 0.2~$M_\odot$ secondary evolving through a CE phase after 0.8 Gyr. We adopted the low-metallicity ($Z=0.004$) initial-to-final mass relations appropriate for the thick disc \citep{rom2015}. The CE phase is followed by cycles of mass transfer and nova ejection of excess material from the binary which, finally, bounces off minimum period leaving it in its present state in $\approx10$~Gyr: $P_{\rm orb}=0.064$~d, $M_{\rm WD}=0.72\,M_\odot$, and $M_{\rm BD}=0.06\,M_\odot$;
The time spent in the CV phase, which maintains a steady white dwarf temperature ($\approx10^4$~K) until it reaches minimum
period, can be as long as $\approx8$~Gyr \citep{bel2018}.
Alternatively, a system with a lower mass primary ($1.5\,M_\odot$) that reaches the CE phase after 2~Gyr could achieve the same state if accreting white dwarfs are allowed to retain some of the transferred mass as proposed by \citet{zor2011}.

\section*{Acknowledgements}

This study is based on observations made with ESO telescopes at the La Silla Paranal Observatory under programme 097.D-0694. We thank Maru\v{s}a \v{Z}erjal for her assistance with the WiFeS pipeline.

\section*{Data Availability}

The FORS2 spectrum is publicly available at the ESO data archive (http://archive.eso.org/cms.html). 
The velocity measurements and the R-band photometric time series are available in electronic form.





\bsp	
\label{lastpage}
\end{document}